\documentclass[letterpaper, 10 pt, conference]{ieeeconf}  

\IEEEoverridecommandlockouts                              
\usepackage{graphicx}
\usepackage{lipsum}  
\usepackage{xcolor}
\usepackage{multirow}
\usepackage{array}
\usepackage{todonotes}
\usepackage{url}

\graphicspath{ {images/} }

\title{\LARGE \bf
CodeVaani: \\ A Multilingual, Voice-Based Code Learning Assistant
}

\author{
Jayant Havare$^{1}$, Srikanth Tamilselvam$^{2}$, Ashish Mittal$^{2}$, Shalaka Thorat$^{1}$, \\
Soham Jadia$^{1}$, Varsha Apte$^{1}$, Ganesh Ramakrishnan$^{1}$\\[1ex]
\small
$^{1}$Indian Institute of Technology Bombay, Mumbai, India\\
$^{2}$IBM Research, India\\
\small
\texttt{\{jayantcse, shalakagt, sohamj, varsha, ganesh\}@cse.iitb.ac.in}\\
\texttt{\{srikanth.tamilselvam, arakeshk\}@in.ibm.com}
}

\begin{document}

\maketitle
\thispagestyle{empty}
\pagestyle{empty}

\begin{abstract}

Programming education often assumes English proficiency and text-based interaction, creating barriers for students from multilingual regions such as India. We present \textit{CodeVaani}, a multilingual speech-driven assistant for understanding code, built into \textit{Bodhitree}~\cite{bodhitree}, a Learning Management System developed at IIT Bombay. It is a voice-enabled assistant that helps learners explore programming concepts in their native languages. The system integrates Indic ASR, a code-aware transcription refinement module, and a code model for generating relevant answers. Responses are provided in both text and audio for natural interaction. In a study with 28 beginner programmers, CodeVaani achieved \textbf{75\% response accuracy}, with over 80\% of participants rating the experience positively. Compared to classroom assistance, our framework offers on-demand availability, scalability to support many learners, and multilingual support that lowers the entry barrier for students with limited English proficiency. The demo will illustrate these capabilities and highlight how voice-based AI systems can make programming education more inclusive.
Supplementary artifacts and demo video are also made available \footnote{\url{https://tinyurl.com/icse2026-artifacts}} \footnote{\url{https://youtu.be/65IpNBB1MDo}}.

\end{abstract}

\section{Introduction}

\begin{figure}
    \centering
    \includegraphics[width=\linewidth]{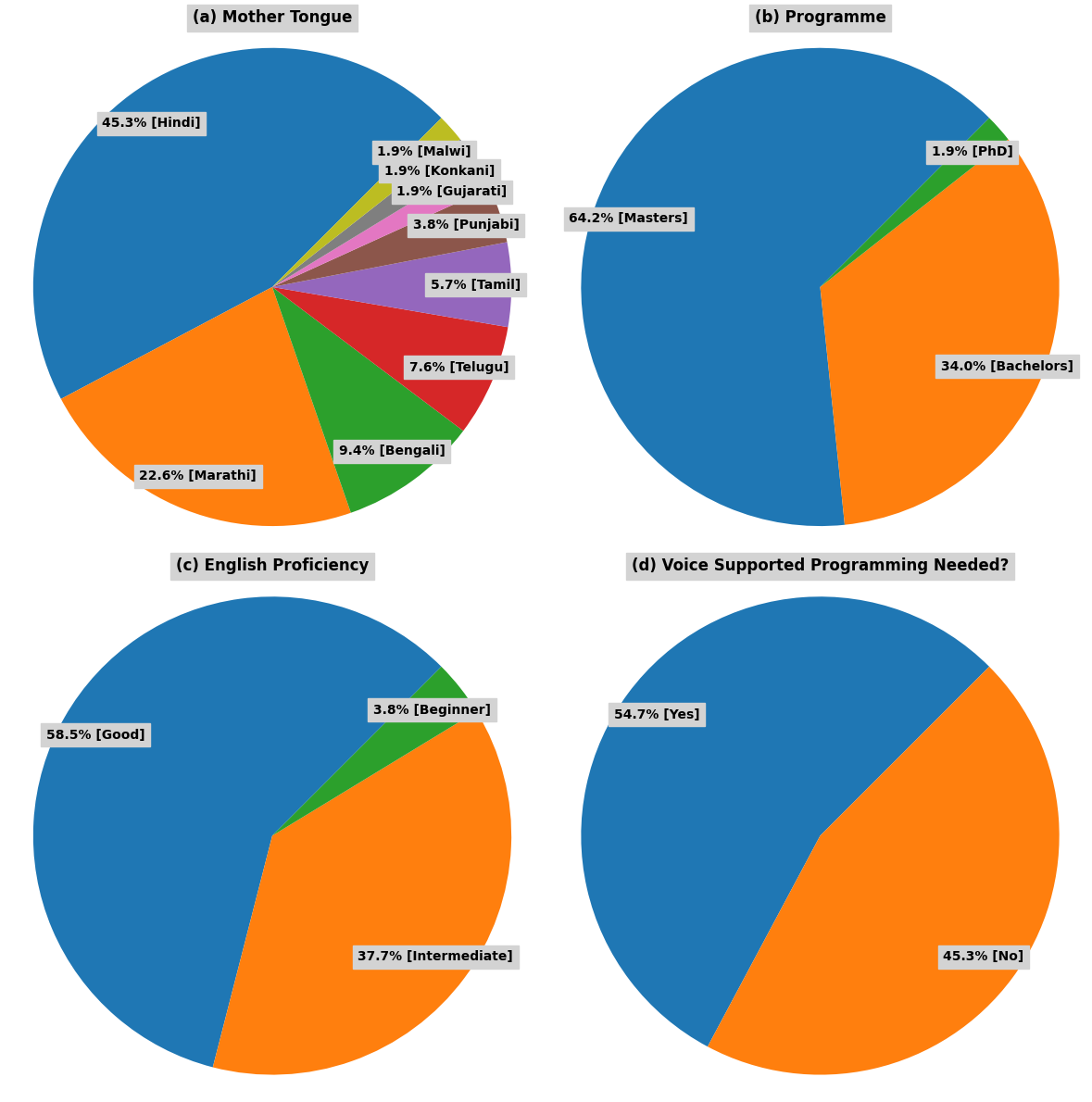}
    \caption{Survey results from 53 engineering students across multiple Indian states, highlighting the multilingual context of programming education.}
    \label{fig:survey}
\end{figure}

The ability to understand the structure, semantics, and behavior of computer programs is a fundamental skill in programming education and practice. With the advent of artificial intelligence (AI), various chat-based assistants have emerged that help a learner interactively understand the code \cite{cursor_ai, visual_studio_code}. However, most of these tools assume English proficiency and rely on text-based, keyboard-driven interaction. This design excludes a large population of learners in multilingual regions such as India, where many students transition directly from regional language schooling to English-medium computer science curricula. For such learners, limited English fluency and typing skills create an additional barrier to mastering programming concepts. Previous work has highlighted the need for inclusive systems that support multilingual learners~\cite{reitmaier2022opportunities, nigatu2024low}, but few tools address this gap in the context of understanding code.

\begin{figure*}[t]
  \centering
  \includegraphics[width=0.8\textwidth]{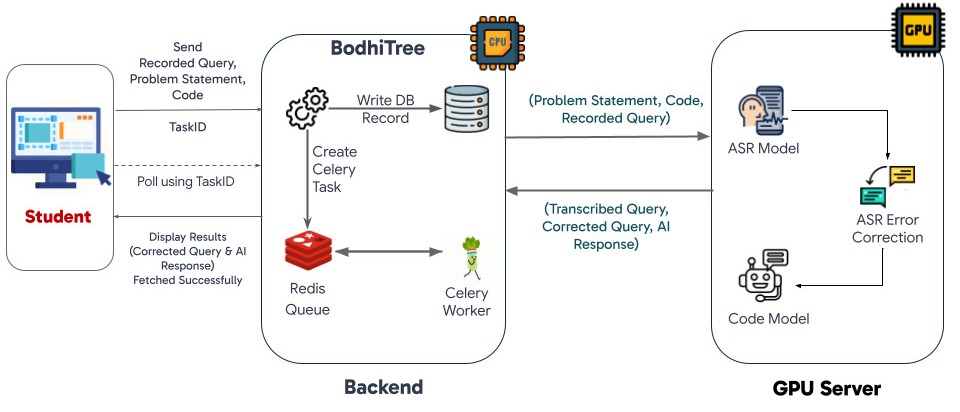}
  \caption{Architecture of CodeVaani: Processing student queries through ASR, error correction, and AI code generation.}
  \label{fig:sys_arch}
\end{figure*}

Voice-based interfaces offer a promising alternative for accessible computing~\cite{paudyal2020voiceye, paudyal2022inclusive}. To better understand the needs of students, we conducted a survey with 53 undergraduate and graduate programmers across multiple Indian states. 
As shown in Figure~\ref{fig:survey}, a majority reported speaking a native language other than English, with only $58.5\%$ considering themselves proficient in English. More than half ($54.7\%$) expressed a preference for voice-based interaction over text, particularly for code-related question answering. These findings highlight an opportunity to reimagine programming support systems that lower the entry barrier through multilingual and multimodal interaction.


However, building such systems is non-trivial. Spoken programming queries are often code-mixed, syntactically irregular, and rich in out-of-vocabulary terms such as identifiers, symbols etc. Standard Automatic Speech Recognition (ASR) systems~\cite{unni2022adaptive, radford2023robust, unni2023improving} frequently misinterpret such inputs, and these errors propagate downstream, significantly degrading the performance of Large Language Models (LLMs) in code understanding tasks~\cite{pan2024lost}.



We present \textit{CodeVaani}, a multilingual speech-driven assistant for understanding code, built into \textit{Bodhitree}~\cite{bodhitree}, a Learning Management System developed at IIT Bombay. It accepts spoken queries in a learner’s native language, transcribes them using ASR, refines the transcription through code-aware LLM-guided processing, and generates relevant answers via a code model. Responses are delivered in both text and audio, enabling accessible, on-demand learning.

\section{CodeVaani: A Speech-Driven Code Assistant}

\begin{figure*}[t]
  \centering
  \includegraphics[width=\textwidth, height=0.3\textheight]{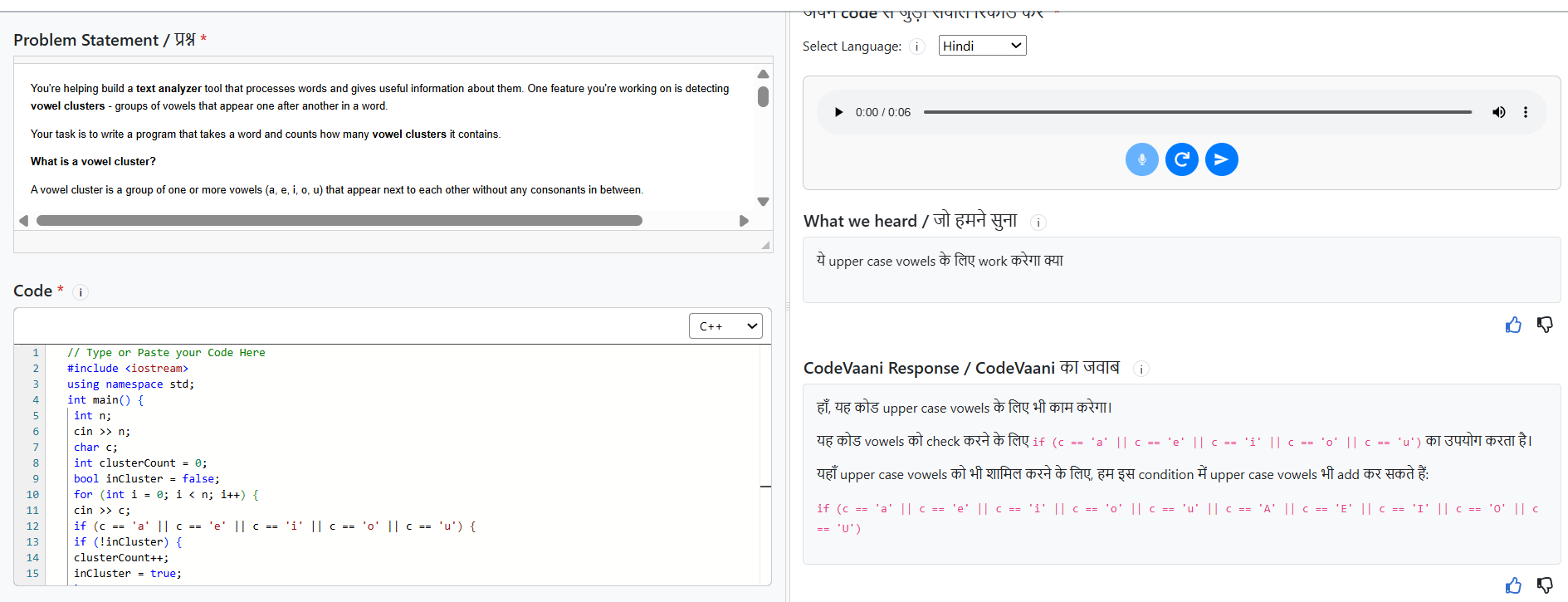}
  \caption{Demo interface where participant’s query was transcribed by ASR, refined by error correction (shown in What we heard field), and answered in the CodeVaani Response field}
  \label{fig:fullwidth}
\end{figure*}

\textit{CodeVaani} enables learners to interact with code using spoken queries in their native language. It is designed to handle multilingual speech, recognize technical terms, and refine transcriptions to ensure reliable code conversations. CodeVaani comprises three main stages: speech recognition, code-aware transcription refinement, and query response generation. Figure~\ref{fig:fullwidth} depicts the demo application interface.

\subsection{Speech Recognition}

\begin{figure}
    \centering
    \includegraphics[width=\linewidth]{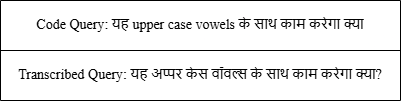}
    \caption{Example of Hindi code-related spoken query and ASR transcriptions. Errors highlight the challenges of code-mixed speech.}
    \label{tab:stt_ex}
\end{figure}

Automatic Speech Recognition (ASR) systems convert speech audio into text, recovering both natural language and code-specific tokens from the audio waveform. Initial experiments with Whisper~\cite{radford2023robust} revealed limitations in handling mixed-language inputs. In particular, Whisper often struggled with queries that interleave native language utterances with English code terms, a common occurrence in student queries. This led to confusions between phonetically similar tokens, especially when identifiers resembled real words (e.g., \texttt{map}, \texttt{sum}). To mitigate this, we employ \texttt{Whisper}~\cite{radford2023robust} for English and \texttt{Indic-Conformer} ASR models from AI4Bharat~\cite{indicconformer} for Indic languages, which are trained on Indian speech data and better preserve mixed-language semantics while improving recognition of technical terms. Fig~\ref{tab:stt_ex} illustrates an example of a spoken Hindi query and its transcription output.

Nevertheless, these models do not fully resolve the problem. We accept the residual ASR errors and address them later through a refinement stage in Section~\ref{subsec:code_aware_refinement}.



\subsection{Code-Aware Transcription Refinement}
\label{subsec:code_aware_refinement}

Despite improvements in ASR, recognition errors persist, particularly for variable names, symbolic operators, and programming keywords embedded within multilingual queries. These errors often stem from phonetic ambiguity or misinterpretation of code tokens as natural language. To address this, we introduce a \emph{code-aware transcription refinement} stage that post-processes ASR outputs to improve fidelity and semantic alignment. We use the instruction-tuned \textbf{gemma-27B}~\cite{team2025gemma} model, further optimized with DPO~\cite{rafailov2023direct}, to correct transcription errors by leveraging the structure and semantics of programming-related language. 
The prompt is designed to: (i) identify and restore misrecognized code terms 
(e.g., variable names, function calls), (ii) correct phonetically distorted 
technical words (e.g., ``ask key'' $\rightarrow$ ``ASCII''), 
(iii) recover symbolic constructs (e.g., \texttt{underscore} $\rightarrow$ \texttt{\_}), 
and (iv) disambiguate between natural and programming language usage.

This strategy allows the model to reconstruct original code-like phrases even when mispronounced or substituted with similar-sounding non-technical words. It is especially valuable in code-mixed and multilingual scenarios where acoustic overlap is common and exact syntax is critical.

\subsection{Query Response Generation}

Finally, the refined transcription is passed as input to the code model, which generates an appropriate response to the user’s query. We use Codestral-22B~\cite{codestral} due to its strong performance on a wide range of coding tasks. In addition, its relatively lightweight architecture makes it well-suited for deployment in classroom or lab environments with limited computational resources.

We tested the model with queries in multiple Indian native languages and observed that it could accurately understand code-related questions and provide contextually relevant responses in the same language. The system also supports a variety of programming tasks, including code explanation, question answering, and basic debugging assistance. By integrating the transcription refinement and code generation stages, our framework ensures that even queries with phonetic errors or code-mixed language are correctly interpreted, resulting in more reliable responses.

\subsection{Implementation Details}
The application components that support CodeVaani (Figure~\ref{fig:sys_arch}) are a ReactJS front end, a Django Python backend, a PostgreSQL database server,  Redis for the task queue, and Celery workers for asynchronous processing.  The AI models used are  Indic-Conformer  and Whisper as the ASR models, gemma-27B for error correction, and the code-specialized LLM, Codestral-22B, for the final response generation. The AI models run on a separate server which has 2 H100 GPUs. We use one GPU for ASR models and an error correction model while another for query response generation. The workflow of CodeVaani is as follows: students record their queries in their preferred language through the browser interface, and submit them. The application additionally sends the problem statement and code and receives a  TaskID in response which is used subsequently to poll for the response. The backend, BodhiTree, logs requests in a database, creates asynchronous Celery tasks, and queues them in Redis, which are then processed by the Celery workers and sent to the GPU server. On the GPU server, the  ASR model transcribes audio, the gemma-27B model corrects and refines the text, and the Codestral-22B model produces the  code responses. The Bodhitree backend collects the corrected query and generated response, delivering them back to the student through the frontend. Overall, the system ensures seamless end-to-end processing from audio query to corrected, AI-generated output.

\section{Evaluation}
\label{evaluation}

To study the effectiveness of our system, we conducted a structured evaluation session in a computer lab. A total of 28 participants, primarily beginners in coding, were invited to interact with the deployed system. Participants were asked to use our system in their mother tongue and record their observations systematically. The study included participants whose native languages were Hindi, Marathi, Gujarati, Tamil, Telugu, Bengali, Malayalam, Kannada, and Odia, representing all major Indian languages. For this purpose, we provided each participant with a journal to document their queries, system responses, and feedback on correctness and appropriateness. Additionally, a follow-up survey was conducted using Google Forms to capture overall usability and satisfaction.

\subsection{Usability Survey}

Participants were asked to rate their overall experience with our system. 
The responses were: Excellent (10.7\%), Good (32.1\%), Satisfactory (28.6\%), 
Fair (10.7\%), and Needs Improvement (17.9\%). These results indicate that more than 80\% of participants rated their experience as satisfactory or above, highlighting the potential of our system as a supportive tool for beginners.

In addition to rating their experience, participants were asked (26 participants were involved): \textit{``If your concerns are resolved, would you like to use our system in your programming course?''} and options: \textit{A. Yes, definitely, B. Yes, probably, C. Maybe, D. Probably not, E. Definitely not}. Responses were as follows: 14 selected A, 11 selected B, 1 selected C, 0 selected D, 0 selected E.  
These results shows participants' strong interest in adopting the system in courses, highlighting its potential as a useful voice-based, multilingual programming aid.

\subsection{Performance}
To quantify the system’s effectiveness, we measured the accuracy of responses as the proportion of correct responses over the total number of queries. 
Participants marked a response as \textit{correct} only if it was exact and fully aligned with their expected answer; partially correct responses were considered incorrect. Based on this strict criterion, the system achieved an accuracy of $75\%$ (72 exact responses out of 96 queries). This performance demonstrates that our system successfully provided correct and relevant responses in 75\% of cases. This reflects a strong foundation for an AI-enabled educational support system.

\subsection{Comparison with Manual Assistance}
In traditional classroom or lab settings, beginners typically rely on teachers or teaching assistants for doubt-solving. While this form of manual assistance is valuable, it comes with several limitations:
\begin{itemize}
    \item Teacher availability is limited to specific days and hours, which restricts continuous support.
    \item A single teacher cannot effectively handle queries from a large number of students simultaneously.
    \item Students do not have the ability to revisit previously asked queries or explanations once the session has ended.
    \item Generally, teachers explain in the English language only. In non-English speaking countries, many students are not so proficient in English.
\end{itemize}

In contrast, our system offers significant advantages:
\begin{itemize}
    \item Availability beyond class hours, enabling on-demand assistance.
    \item Scalability to support many students simultaneously without additional human effort.
    \item Consistent responses that can be revisited, allowing learners to reinforce concepts at their own comfort.
    \item Multilingual support, which bridges the gap for students who find it easier to learn in their mother tongue.
\end{itemize}

This comparison highlights that while manual assistance is essential for personalized mentoring, \textit{CodeVaani} complements it by addressing scalability, availability, and multilingual accessibility—factors that traditional classroom assistance alone cannot meet.



\section{Conclusion \& Future Work}
\label{conclusion}


In this paper, we presented a speech-driven multilingual code assistant that enables beginner programmers to reason code concepts in their native languages. Our system integrates ASR models for Indic languages, a code-aware transcription refinement stage, and a code model to generate relevant responses. Evaluation results show 75\% accuracy and positive user feedback, highlighting its potential as an educational support tool.  This work demonstrates that voice-based interfaces can significantly lower entry barriers for students who struggle with English proficiency and keyboard-driven interaction. 

Future work will extend CodeVaani in the following  directions: (i) enabling multi-turn conversational support for interactive dialogue, (ii) fine-tuning ASR and transcription models on larger code-mixed speech datasets to improve real-world accuracy, and (iii) merging ASR error correction and response generation into a unified pipeline to reduce latency. Furthermore, we aim to support some desired features mentioned in the survey, such as typed input in multiple languages. 



\medskip
 
\bibliographystyle{unsrt}  
\bibliography{sample-base}

\end{document}